\documentclass[aps,preprint]{revtex4}%
\usepackage[USenglish]{babel}
\usepackage{amsmath}
\usepackage{graphicx}
\usepackage{amsfonts}
\usepackage{eurosym}
\usepackage{amssymb}
\usepackage{float}
\usepackage{hyphenat}%
\providecommand{\U}[1]{\protect \rule{.1in}{.1in}}
\providecommand{\U}[1]{\protect \rule{.1in}{.1in}}

\begin{document}
\title{Path integral approach to Asian options in the Black-Scholes model}
\author{J.P.A. Devreese$^{1}$}
\author{D. Lemmens$^{1}$}
\author{J. Tempere$^{1,2}$}
\affiliation{$^{1}$ TQC, Universiteit Antwerpen, Groenenborgerlaan 171, B2020 Antwerpen,\ Belgium}
\affiliation{$^{2}$ Lyman Laboratory of Physics, Harvard University, Cambridge MA02138, USA}

\begin{abstract}
We derive a closed-form solution for the price of an average price as well as
an average strike geometric Asian option, by making use of the path integral
formulation. Our results are compared to a numerical Monte Carlo simulation.
We also develop a pricing formula for an Asian option with a barrier on a
control process, combining the method of images with a partitioning of the set
of paths according to the average along the path. This formula is exact when
the correlation is zero, and is approximate when the correlation increases.

\end{abstract}
\date{June 2, 2009}
\maketitle

\section{Introduction}

Since the beginning of financial science, stock prices, option prices and
other quantities have been described by stochastic and partial differential
equations. Since the 1980s however, the path integral approach, created in the
context of quantum mechanics by Richard Feynman \cite{Feynman}, has been
introduced to the field of finance \cite{Dash 1,Dash 2}. Earlier, Norbert
Wiener \cite{Wiener}, in his studies on Brownian motion and the Langevin
equation, used a type of functional integral that turns out to be a special
case of the Feynman path integral (see also Mark Kac \cite{Kac}, and for a
general overview see Kleinert \cite{Kleinert} and Schulman \cite{Schulman}).
The power of path-integration for finance (\cite{Kleinert}%
,\cite{Rosa-Clot,Rosa-Clot 2,Montagna,Bormetti,Kleinert 2,Baaquie,Bouchaud}%
)\ lies in its ability to naturally account for payoffs that are
path-dependent. This makes path integration the method of choice to treat one
of the most challenging types of derivatives, the path-dependent options.
Feynman and Kleinert \cite{Feynman-Kleinert} showed how quantum-mechanical
partition functions can be approximated by an effective classical partition
function, a technique which has been successfully applied to the pricing of
path-dependent options (see ref. \cite{Kleinert} and references therein, and
Refs. \cite{Kleinert 2,Damiaan} for recent applications).

There exist many different types of path-dependent options. The two types
which are considered in this paper are Asian and barrier options. Asian
options are exotic path-dependent options for which the payoff depends on the
average price of the underlying asset during the lifetime of the option
\cite{Hull,Bormetti,Kemna Vorst,Turnbull Wakeman}. One distinguishes between
\textit{average price} and \textit{average strike} Asian options. The average
price Asian option has been treated in the context of path integrals by
Linetsky \cite{Linetsky}. The payoff of an average price is given by
$\max(\bar{S}_{T}-K,0)$ and $\max(K-\bar{S}_{T},0)$ for a call and put option
respectively. Here $K$ is the strike price and $\bar{S}_{T}$ denotes the
average price of the underlying asset at maturity $T$. $\bar{S}_{T}$ can
either be the arithmetical or geometrical average of the asset price. Average
price Asian options cost less than plain vanilla options. They are useful in
protecting the owner from sudden short-lasting price changes in the market,
for example due to order imbalances \cite{Rosenow}. Average strike options are
characterized by the following payoffs: $\max(S_{T}-\bar{S}_{T},0)$ and
$\max(\bar{S}_{T}-S_{T},0)$ for a call and put option respectively, where
$S_{T}$ is the price of the underlying asset at maturity $T$. Barrier options
are options with an extra boundary condition. If the asset price of such an
option reaches the barrier during the lifetime of the option, the option
becomes worthless, otherwise the option has the same payoff as the option on
which the barrier has been imposed. (for more information on exit-time
problems see Ref. \cite{Masilover} and the references therein)

In section \ref{average strike option} we treat the geometrically averaged
Asian option. In section \ref{1} the asset price propagator for this standard
Asian option is derived within the path integral framework in a similar
fashion as in Ref. \cite{Linetsky} for the weighted Asian option. The
underlying principle of this derivation is the effective classical partition
function technique developed by Feynman and Kleinert \cite{Feynman-Kleinert}.
In section \ref{2}\ we present an alternative derivation of this propagator
using a stochastic calculus approach. This propagator now allows us to price
both the average price and average strike Asian option. For both types of
options this results in a pricing formula which is of the same form as the
Black-Scholes formula for the plain vanilla option. Our result for the option
price of an average price Asian option confirms the result found in the
literature \cite{Linetsky,Lipton}. For the average strike option no formula of
this simplicity exists as far as we know. Our derivation and analysis of this
formula is presented in section \ref{3}, where our result is checked with a
Monte Carlo simulation. In section \ref{4}\ we impose a boundary condition on
the Asian option in the form of a barrier on a control process, and check
whether the method used in section \ref{average strike option} is still valid
when this boundary condition is imposed on the propagator for the normal Asian
option, using the method of images. Finally in Section \ref{5} we draw conclusions.

\section{Geometric Asian options in the Black-Scholes
model\label{average strike option}}

\subsection{Partitioning the set of all paths\label{1}}

The path integral propagator is used in financial science to track the
probability distribution of the logreturn $x_{t}=\log(S_{t}/S_{0})$ at time
$t$, where $S_{0}$ is the initial value of the underlying asset. This
propagator is calculated as a weighted sum over all paths from the initial
value $x_{0}=0$ at time $t=0$ to a final value $x_{T}=\log(S_{T}/S_{0})$ at
time $t=T:$%
\begin{equation}
\mathcal{K}\left(  x_{T},T|0,0\right)  =\int \mathcal{D}x\text{ }\exp \left(
-\int_{0}^{T}\mathcal{L}_{BS}\left[  x(t)\right]  \,dt\right)
\end{equation}
The weight of a path, in the Black-Scholes model, is determined by the
Lagrangian%
\begin{equation}
\mathcal{L}_{BS}\left[  x(t)\right]  =\frac{1}{2\sigma^{2}}\left[  \dot
{x}-\left(  \mu-\frac{\sigma^{2}}{2}\right)  \right]  ^{2}
\label{Black-Scholes Lagrangiaan}%
\end{equation}
where $\mu$ is the drift and $\sigma$ is the volatility appearing in the
Wiener process for the logreturn \cite{Rosa-Clot}.

For Asian options, the payoff is a function of the average value of the asset.
Therefore we introduce $\bar{x}_{T}=\log \left(  \bar{S}_{T}/S_{0}\right)  $ as
the logreturn corresponding to the average asset price at maturity $T$. When
$\bar{S}_{T}$ is the geometric average of the asset price, then $\bar{x}_{T}$
is an algebraic average.
\begin{equation}
\bar{x}_{T}=\frac{1}{T}\int_{0}^{T}x(t)dt. \label{gemiddelde}%
\end{equation}
The key step to treat Asian options within the path integral framework is to
partition the set of all paths into subsets of paths, where each path in a
given subset has the same average $\bar{x}_{T}$. Summing over only these paths
that have a given average $\bar{x}_{T}$ defines the conditional propagator
$\mathcal{K}\left(  x_{T},T\, \left \vert 0,0\right \vert \bar{x}_{T}\right)
$:
\begin{equation}
\mathcal{K}\left(  x_{T},T\, \left \vert 0,0\right \vert \bar{x}_{T}\right)
=\int \mathcal{D}x\, \delta \left(  \bar{x}_{T}-\frac{1}{T}\int_{0}%
^{T}x(t)\,dt\right)  \exp \left(  -\int_{0}^{T}\mathcal{L}_{BS}\left[
x(t)\right]  \,dt\right)  \label{Conditionele propagator}%
\end{equation}
This is indeed a partitioning of the sum over all paths:%
\begin{equation}
\mathcal{K}\left(  x_{T},T|0,0\right)  =%
{\displaystyle \int \limits_{-\infty}^{\infty}}
d\bar{x}_{T}\text{ }\mathcal{K}\left(  x_{T},T\, \left \vert 0,0\right \vert
\bar{x}_{T}\right)
\end{equation}
The delta function in the sum $\int \mathcal{D}x$ over all paths picks out
precisely all the paths that will have the same payoff for an Asian option.

The calculation of $\mathcal{K}\left(  x_{T},T\, \left \vert 0,0\right \vert
\bar{x}_{T}\right)  $ is straightforward; when the delta function is rewritten
as an exponential,
\begin{equation}
\mathcal{K}\left(  x_{T},T\, \left \vert 0,0\right \vert \bar{x}_{T}\right)  =%
{\displaystyle \int \limits_{-\infty}^{\infty}}
\frac{dk}{2\pi}e^{ikx}\int \mathcal{D}x\, \exp \left(  -\int_{0}^{T}\left(
\mathcal{L}_{BS}\left[  x(t)\right]  +\frac{1}{T}ikx(t)\right)  dt\right)  ,
\end{equation}
the resulting Lagrangian is that of a free particle in a constant force field
in 1D. The resulting integration over paths is found by standard procedures
\cite{Feynman Hibbs}:
\begin{align}
\mathcal{K}\left(  x_{T},T\, \left \vert 0,0\right \vert \bar{x}_{T}\right)   &
=\frac{\sqrt{3}}{\pi \sigma^{2}T}\exp \left \{  -\frac{1}{2\sigma^{2}T}\left[
x_{T}-\left(  \mu-\frac{\sigma^{2}}{2}\right)  T\right]  ^{2}\right.
\nonumber \\
&  \left.  -\frac{6}{\sigma^{2}T}\left(  \bar{x}_{T}-\frac{x_{T}}{2}\right)
^{2}\right \}  , \label{Conditionele propagator uitgerekend}%
\end{align}
and corresponds to the result found by Kleinert \cite{Kleinert} and by
Linetsky \cite{Linetsky}.

\bigskip

\subsection{Link with stochastic calculus\label{2}}

The conditional propagator $\mathcal{K}\left(  x_{T},T\, \left \vert
0,0\right \vert \bar{x}_{T}\right)  $ is interpreted in the framework of
stochastic calculus as the joint propagator $\mathcal{K}\left(  x_{T},\bar
{x}_{T},T\,|0,0,0\right)  $ of $x_{T}$ and its average $\bar{x}_{T}$. The
calculation of $\mathcal{K}\left(  x_{T},\bar{x}_{T},T\,|0,0,0\right)  $ here
is similar to the derivation presented in Ref. \cite{Glassy}\ where this joint
propagator is calculated for the Vasicek model. The main point is that in a
Gaussian model the joint distribution of the couple $\left \{  x_{T},\bar
{x}_{T}\right \}  $ has to be Gaussian too. As a consequence this joint
distribution is fully characterized by the expectation values and the
variances of $x_{T}$ and $\bar{x}_{T}$\ and by the correlation between these
two processes. The expectation value of $\bar{x}_{T}\left(  t\right)  $ is
given by $\left(  \mu-\frac{\sigma^{2}}{2}\right)  \frac{t}{2}$, its variance
by $\frac{\sigma^{2}t}{3}$ and the correlation between the two processes by
$\frac{\sqrt{3}}{2}$. The density function of such a Gaussian process is then
known to be%
\begin{align}
\mathcal{K}\left(  x_{T},\bar{x}_{T},T\,|0,0,0\right)   &  =\frac{\sqrt{3}%
}{\pi \sigma^{2}T}\exp \left(  -\frac{2}{\sigma^{2}T}\left \{  \left[
x_{T}-\left(  \mu-\frac{\sigma^{2}}{2}\right)  T\right]  ^{2}+3\left[  \bar
{x}_{T}-\left(  \mu-\frac{\sigma^{2}}{2}\right)  \frac{T}{2}\right]
^{2}\right.  \right. \nonumber \\
&  \left.  \left.  -3\left[  x_{T}-\left(  \mu-\frac{\sigma^{2}}{2}\right)
T\right]  \left[  \bar{x}_{T}-\left(  \mu-\frac{\sigma^{2}}{2}\right)
\frac{T}{2}\right]  \right \}  \right)
\end{align}
This agrees with Eq. (\ref{Conditionele propagator uitgerekend}) for
$\mathcal{K}\left(  x_{T},T\, \left \vert 0,0\right \vert \bar{x}_{T}\right)
$.\bigskip

\subsection{Pricing of an average strike geometric Asian option\label{3}}

If the payoff at time $T$ of an Asian option is written as $V_{T}%
^{Asian}(x_{T},\bar{x}_{T})$, then the expected payoff is%
\begin{equation}
\mathbb{E}\left[  V_{T}^{Asian}(x_{T},\bar{x}_{T})\right]  =%
{\displaystyle \int \limits_{-\infty}^{\infty}}
dx_{T}%
{\displaystyle \int \limits_{-\infty}^{\infty}}
d\bar{x}_{T}\text{ }V_{T}^{Asian}(x_{T},\bar{x}_{T})\mathcal{K}\left(
x_{T},T\, \left \vert 0,0\right \vert \bar{x}_{T}\right)
\label{algemene vorm waarde van de aziatische optie}%
\end{equation}
The price of the option, $V_{0}^{Asian}$ is the discounted expected payoff,
\begin{equation}
V_{0}^{Asian}=e^{-rT}\mathbb{E}\left[  V_{T}^{Asian}(x_{T},\bar{x}%
_{T})\right]  \label{6}%
\end{equation}
where $r$ is the discount (risk-free) interest rate. Using expression
(\ref{algemene vorm waarde van de aziatische optie}) the price of any option
which is dependent on the average of the underlying asset during the lifetime
of the option can be calculated. We will now derive the price of an average
strike geometric Asian call option explicitly. In order to do this, expression
(\ref{algemene vorm waarde van de aziatische optie}) has to be evaluated using
the payoff:
\begin{equation}
V_{T}^{Asian}(x_{T},\bar{x}_{T})=\max(S_{T}-\bar{S}_{T},0)=S_{0}\max(e^{x_{T}%
}-e^{\bar{x}_{T}},0) \label{payoff}%
\end{equation}
Substituting (\ref{payoff}) in (\ref{6}) yields%
\begin{equation}
V_{0}^{Asian}=S_{0}e^{-rT}%
{\displaystyle \int \limits_{-\infty}^{\infty}}
d\bar{x}_{T}%
{\displaystyle \int \limits_{\bar{x}_{T}}^{\infty}}
dx_{T}\text{ }\left(  e^{x_{T}}-e^{\bar{x}_{T}}\right)  \mathcal{K}\left(
x_{T},T\, \left \vert 0,0\right \vert \bar{x}_{T}\right)
\end{equation}
where the lower boundary of the $x_{T}$ integration now depends on $\bar
{x}_{T}$. When considering an average price call, the payoff (for a call
option) is $\max(\bar{S}_{T}-K,0)$ leading to a constant lower boundary
$\log(K/S_{0})$ for the $\bar{x}_{T}$ integration, and the integrals are
easily evaluated. In the present case however, the integration boundary is
more complicated and it is more convenient to express this boundary through a
Heaviside function, written in its integral representation:
\begin{align}
V_{0}^{Asian}  &  =S_{0}\,e^{-rT}\, \frac{1}{2\pi i}\int_{-\infty}^{+\infty
}d\bar{x}_{T}\, \, \int_{-\infty}^{+\infty}dx_{T}\int_{-\infty}^{+\infty}%
d\tau \, \frac{e^{i\left(  x_{T}-\bar{x}_{T}\right)  \tau}}{\tau-i\varepsilon
}\label{Waarde van de Aziatische optie 1}\\
&  \times \left(  e^{x_{T}}-e^{\bar{x}_{T}}\right)  \mathcal{K}\left(
x_{T},T\, \left \vert 0,0\right \vert \bar{x}_{T}\right) \nonumber
\end{align}
Now the two original integrals have been reduced to Gaussians at the cost of
inserting a complex term in the exponential. Expression
(\ref{Waarde van de Aziatische optie 1}) can be split into two terms denoted
$I_{1}$ and $I_{2}$, where%
\begin{align}
I_{1}  &  =S_{0}e^{-rT}\frac{\sqrt{3}}{\pi \sigma^{2}T}\, \frac{1}{2\pi i}%
\int_{-\infty}^{+\infty}d\tau \frac{1}{\tau-i\varepsilon}\int_{-\infty
}^{+\infty}d\bar{x}_{T}\int_{-\infty}^{+\infty}dx_{T}\nonumber \\
&  \times \exp \left \{  -\frac{1}{2\sigma^{2}T}\left[  x_{T}-\left(  \mu
-\frac{\sigma^{2}}{2}\right)  T\right]  ^{2}\right. \nonumber \\
&  \left.  -\frac{6}{\sigma^{2}T}\left(  \bar{x}_{T}-\frac{x_{T}}{2}\right)
^{2}+i\left(  x_{T}-\bar{x}_{T}\right)  \tau+x_{T}\right \}
\end{align}
and $I_{2}$ has the same form, except with $\bar{x}_{T}$ instead of $x_{T}$ in
the last term of the argument of the exponent. As a first step, the Gaussian
integrals over $x_{T}$ and $\bar{x}_{T}$ are calculated, yielding%
\begin{equation}
I_{1}=S_{0}\,e^{-\left(  r-\mu \right)  T}\frac{1}{2\pi i}\int_{-\infty
}^{+\infty}\frac{f\left(  \tau \right)  }{\tau-i\varepsilon}d\tau \label{I 3}%
\end{equation}
with%
\begin{equation}
f\left(  \tau \right)  =\exp \left[  -\frac{\sigma^{2}T}{6}\tau^{2}+\left(
\mu+\frac{\sigma^{2}}{2}\right)  \frac{iT}{2}\tau \right]
\end{equation}
Now the integral has been reduced to a form which can be rewritten by making
use of Plemelj's formulae. Taking into account symmetry, this reduces to%
\begin{equation}
\int_{-\infty}^{+\infty}\frac{f\left(  \tau \right)  }{\tau-i\varepsilon}%
d\tau=i\pi \left[  \operatorname{erf}\left(  \frac{b}{2\sqrt{a}}\right)
+1\right]
\end{equation}
with%
\begin{equation}
\left \{
\begin{array}
[c]{l}%
\smallskip a=\dfrac{\sigma^{2}T}{6}\\
\smallskip b=\left(  \mu+\dfrac{\sigma^{2}}{2}\right)  \dfrac{T}{2}%
\end{array}
\right.
\end{equation}
The first term thus becomes%
\begin{equation}
I_{1}=S_{0}e^{-rT}\frac{\sqrt{3}}{\pi \sigma^{2}T}\, \frac{1}{2}\left \{
\operatorname{erf}\left[  \sqrt{\frac{3T}{8\sigma^{2}}}\left(  \mu
+\frac{\sigma^{2}}{2}\right)  \right]  +1\right \}
\end{equation}
The second term, $I_{2},$ is evaluated similarly, leading to
\begin{align}
V_{0}^{Asian}  &  =S_{0}e^{-rT}\left(  \frac{\sqrt{3}}{\pi \sigma^{2}T}\,
\frac{1}{2}\left \{  \operatorname{erf}\left[  \sqrt{\frac{3T}{8\sigma^{2}}%
}\left(  \mu+\frac{\sigma^{2}}{2}\right)  \right]  +1\right \}  \right.
\nonumber \\
&  \left.  -\exp \left[  \left(  \mu-\frac{\sigma^{2}}{6}\right)  \frac{T}%
{2}\right]  \left \{  \operatorname{erf}\left[  \sqrt{\frac{3T}{8\sigma^{2}}%
}\left(  \mu-\frac{\sigma^{2}}{6}\right)  \right]  +1\right \}  \right)
\end{align}
Using the cumulative distribution function of the normal distribution \-%
\begin{equation}
\Phi \left(  x\right)  =\nolinebreak \frac{1}{2}\left[  1+\operatorname{erf}%
\left(  \frac{x}{\sqrt{2}}\right)  \right]
\end{equation}
this can be rewritten in a more compact form as%
\begin{equation}
V_{0}^{Asian}=S_{0}e^{-rT}\left(  e^{\mu T}\, \, \Phi \left(  d_{1}\right)
-e^{\left(  \mu-\frac{\sigma^{2}}{6}\right)  \frac{T}{2}}\Phi \left(
d_{2}\right)  \right)  \label{priceform}%
\end{equation}
with the following shorthand notations%
\begin{equation}
\left \{
\begin{array}
[c]{c}%
\smallskip d_{1}=\sqrt{\dfrac{3T}{4\sigma^{2}}}\left(  \mu+\dfrac{\sigma^{2}%
}{2}\right) \\
\smallskip d_{2}=\sqrt{\dfrac{3T}{4\sigma^{2}}}\left(  \mu-\dfrac{\sigma^{2}%
}{6}\right)
\end{array}
\right.
\end{equation}
Expression (\ref{priceform}) is the analytic pricing formula for an average
strike geometric Asian call option, obtained in the present work with the path
integral formalism. To the best of our knowledge, no pricing formula of this
simplicity exists. To check this formula, we compared its results to those of
a Monte Carlo simulation. The Monte Carlo scheme used is as follows
\cite{Glassy}: first, the evolution of the logreturn is simulated for a large
number of paths. This evolution is governed by a discrete geometric Brownian
motion for a number of time steps. Using the value for the logreturn at each
time step, the average logreturn can be calculated for every path.
Subsequently the payoff per path can be obtained, which is then used to
calculate the option price by averaging over all payoffs per path en
discounting back in time. The analytical result and the Monte Carlo simulation
agree to within a relative error of 0.3\% when 500 000 samples and 100 time
steps are used. This means that our analytical result lies within the error
bars at every point. We also obtained the result for an average price Asian
option; in contrast to the new result for the average strike option this could
be compared to the existing formula \cite{Linetsky,Lipton}, and was found to
be the same.

\section{Asian option with a barrier on a control process\label{4}}

\subsection{Derivation of the option
price\label{derivation of the option price}}

\begin{figure}
[h]
\begin{center}
\includegraphics[
height=5.1026in,
width=4.0965in
]%
{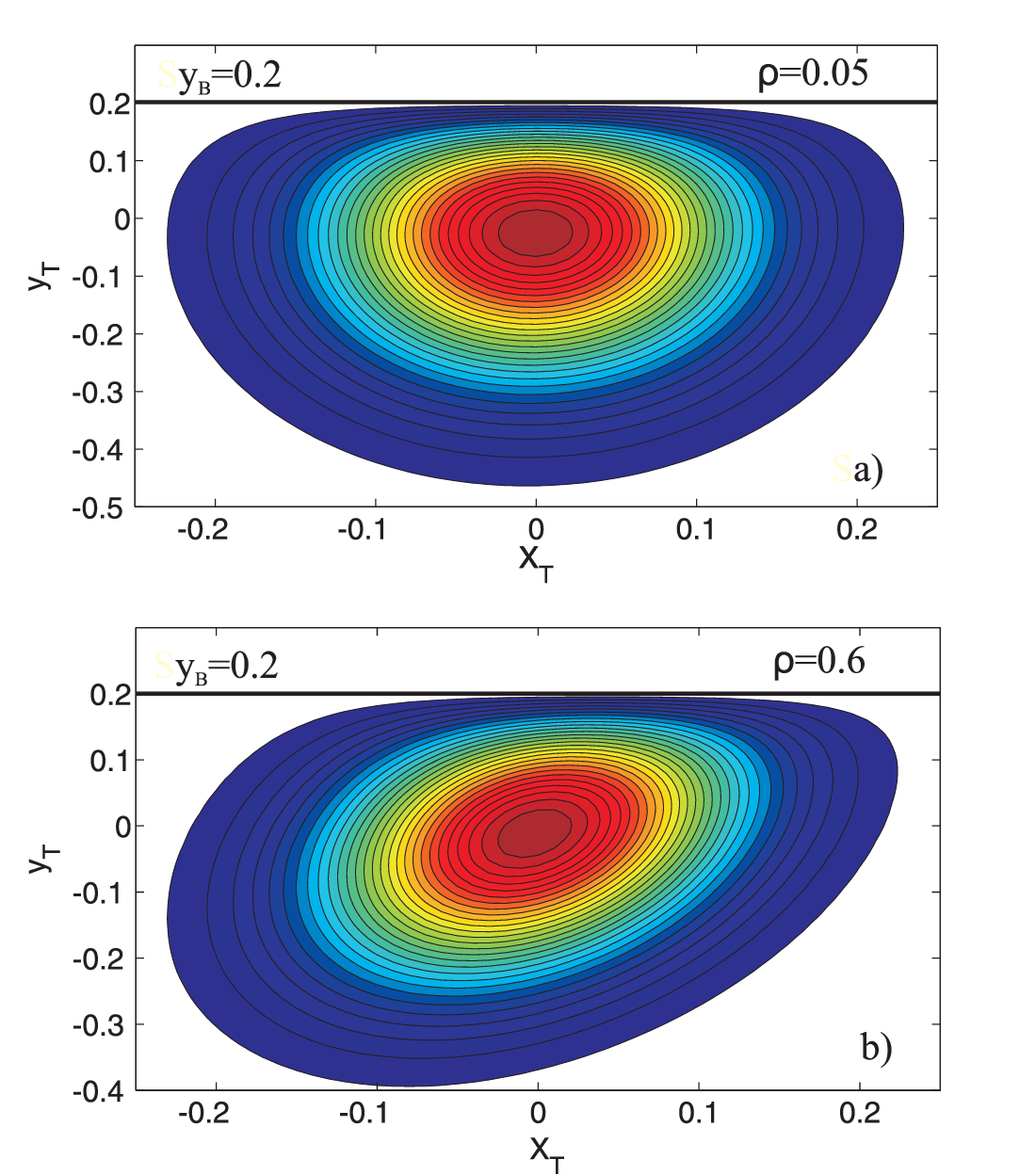}%
\caption{Graphical representation of the barrier-propagator
(\ref{propagator voor aziaat met barriere}) in arbitrary units, for the system
of two correlated processes $x$ and $y$, given by (\ref{x,y}), where a barrier
has been placed on the $y$ process at $y_{B}=0.2$. The following values were
used in this figure: $\mu=\nu=0.03\ $Year$^{-1};\sigma=\xi=0.25~$%
Year$^{-\frac{1}{2}};T=1~$Year}%
\label{propagator_controle_bijeen}%
\end{center}
\end{figure}

In this case we consider two stochastic processes:
\begin{equation}
\left \{
\begin{array}
[c]{l}%
\smallskip dx=\left(  \mu-\dfrac{\sigma^{2}}{2}\right)  dt+\sigma dW_{1}\\
\smallskip dy=\left(  \nu-\dfrac{\xi^{2}}{2}\right)  dt+\xi dZ
\end{array}
\right.  \label{x,y}%
\end{equation}
which are correlated in the following manner: $\left \langle dW_{1}%
dZ\right \rangle =\rho dt$.~The $x$ process models the logreturn of the asset
price which underlies the Asian option, and the $y$ process describes the
control process. The payoff for an Asian option with a barrier on a control
process is the same as for a normal Asian option, with the extra condition
that the payoff is zero whenever the value of $\allowbreak y$ surpasses a
certain predetermined barrier. This is an example of an \textit{up-and-out
barrier}. There are other types of barrier options, namely down-and-out etc.,
but since their treatment is analogous we will not consider them here. The
payoff for an Asian option with a barrier on a control process is given by:
\begin{equation}
V_{T}^{AB}(x_{T},\bar{x}_{T})=\left \{
\begin{array}
[c]{l}%
S_{0x}e^{\bar{x}_{T}}-K\text{ \  \  \  \ }\forall t\in \left[  0,T\right]
:y\left(  t\right)  <y_{B}\\
~\  \ 0\text{ \ ~~~~\  \ ~~~\ ~~~~}\exists t\in \left[  0,T\right]  :y\left(
t\right)  \geq y_{B}%
\end{array}
\right.  \label{payoff2}%
\end{equation}
where the payoff of an average price Asian option has been used. Here $S_{0x}$
denotes the initial asset price of the asset corresponding to the logreturn
$x$ and $y_{B}$ is the value of the barrier which has been placed upon the $y$
process. It is difficult to price this option using payoff (\ref{payoff2})
because of the extra barrier condition. However, if this condition could
somehow be included in the propagator for these two processes, then the payoff
would reduce to that of a normal (average price) Asian option, making the
calculations more tractable. To construct this new propagator, henceforth
called \textit{barrier-propagator}, a linear combination of propagators for
the combined evolution of both processes given in (\ref{x,y}) can be taken:
\begin{equation}
\mathcal{K}_{y}^{B}\left(  x_{T},y_{T},\bar{x}_{T},T\,|0,0,0,0\right)
=\mathcal{K}\left(  x_{T},y_{T},\bar{x}_{T},T\,|0,0,0,0\right)  +C~\mathcal{K}%
\left(  x_{T},y_{T},\bar{x}_{T},T\,|x_{S},y_{S},\bar{x}_{S},0\right)
\end{equation}
where $\mathcal{K}_{y}^{B}$ stands for the propagator for the processes $x$
and $y$ where a barrier condition has been placed upon the $y$ process. The
propagator $\mathcal{K}\left(  x_{T},y_{T},\bar{x}_{T},T\,|0,0,0,0\right)  $
belonging to the system (\ref{x,y}) is an extension of the propagator
(\ref{Conditionele propagator uitgerekend}), and is given by:%
\begin{align}
\mathcal{K}\left(  x_{T},y_{T},\bar{x}_{T},T\,|0,0,0,0\right)   &
=\sqrt{\frac{3}{2\pi^{3}T^{3}\sigma^{4}\xi^{2}\left(  1-\rho^{2}\right)  }%
}\nonumber \\
&  \times \exp \left \{  \frac{\rho}{\sigma \xi \left(  1-\rho^{2}\right)
T}\left[  x_{T}-\left(  \mu-\frac{\sigma^{2}}{2}\right)  T\right]  \left[
y_{T}-\left(  \nu-\frac{\xi^{2}}{2}\right)  T\right]  \right. \nonumber \\
&  -\frac{1}{2\sigma^{2}\left(  1-\rho^{2}\right)  T}\left[  x_{T}-\left(
\mu-\frac{\sigma^{2}}{2}\right)  T\right]  ^{2}-\frac{1}{2\xi^{2}\left(
1-\rho^{2}\right)  T}\nonumber \\
&  \left.  \times \left[  y_{T}-\left(  \nu-\frac{\xi^{2}}{2}\right)  T\right]
^{2}-\frac{6}{\sigma^{2}T}\left(  \bar{x}_{T}-\frac{x_{T}}{2}\right)
^{2}\right \}  \label{Propagator controleproces}%
\end{align}
Furthermore C is a factor upon which three conditions will be placed and
$x_{S},y_{S}$ represent the initial condition from which the mirror-propagator
starts. This mirror-propagator is used to eliminate all paths that cross the
barrier, and because the paths represented by the mirror-propagator usually
have higher values than the paths represented by the propagator $\mathcal{K}%
\left(  x_{T},y_{T},\bar{x}_{T},T\,|0,0,0,0\right)  $, they have been given
another average $\bar{x}_{S}$. The barrier-propagator must be zero at the
boundary:%
\begin{equation}
\mathcal{K}_{y}^{B}\left(  x_{T},y_{B},\bar{x}_{T},T\,|0,0,0,0\right)  =0
\end{equation}
Using this boundary condition, an expression for C can be derived which must
satisfy three conditions: firstly C must be independent of the averages
$\bar{x}_{T}$ and $\bar{x}_{S}$, secondly it may not depend on $x_{T}$ and
finally it must be time-independent. This eventually leads to the following
propagator for the total system of correlated stochastic processes $x$ and
$y$, with a barrier condition on $y$ when $y_{T}\in \left[  -\infty
,y_{B}\right[  $:
\begin{align}
\mathcal{K}_{y}^{B}\left(  x_{T},y_{T},\bar{x}_{T},T\,|0,0,0,0\right)   &
=\mathcal{K}\left(  x_{T},y_{T},\bar{x}_{T},T\,|0,0,0,0\right) \nonumber \\
&  -e^{\frac{2y_{B}}{\xi \left(  4-3\rho^{2}\right)  }\left[  \frac{4}{\xi
}\left(  \nu-\frac{\xi^{2}}{2}\right)  -3\frac{\rho}{\sigma}\left(  \mu
-\frac{\sigma^{2}}{2}\right)  \right]  }~\mathcal{K}\left(  x_{S},y_{S}%
,\bar{x}_{S},T\,|0,0,0,0\right)  \label{propagator voor aziaat met barriere}%
\end{align}
with the following shorthand notations:%
\begin{equation}
\left \{
\begin{array}
[c]{l}%
x_{S}=\dfrac{2y_{B}}{\xi}\allowbreak \dfrac{\rho \sigma}{\left(  4-3\rho
^{2}\right)  }\\
\bar{x}_{S}=\dfrac{-\left(  x_{S}-x_{T}\right)  +\sqrt{\left(  x_{S}%
-x_{T}\right)  ^{2}+4~\bar{x}_{T}\left(  \bar{x}_{T}-x_{T}\right)  }}{2}\\
y_{S}=2y_{B}%
\end{array}
\right.  \medskip \label{xbars}%
\end{equation}
The propagator (\ref{propagator voor aziaat met barriere}) is equal to zero
when $y_{T}\in \left[  y_{B},+\infty \right]  $. A graphical presentation of
propagator (\ref{propagator voor aziaat met barriere}) is shown in Fig.
\ref{propagator_controle_bijeen}.%

Using the propagator (\ref{propagator voor aziaat met barriere}) the price of
an Asian option with a barrier $V_{0}^{AB}$ can be calculated. The general
pricing formula is given by:%
\begin{equation}
V_{0}^{AB}=e^{-rT}\int_{-\infty}^{y_{B}}dy_{T}\, \int_{-\infty}^{+\infty}%
d\bar{x}_{T}\, \, \int_{-\infty}^{+\infty}dx_{T}\, \mathcal{K}_{y}^{B}\left(
x_{T},y_{T},\bar{x}_{T},T\,|0,0,0,0\right)  \,V_{T}^{Asian}(x_{T},\bar{x}_{T})
\label{algemene prijs formule aziaat barriere}%
\end{equation}
This calculation was done for an average price option: $V_{T}^{Asian}%
(x_{T},\bar{x}_{T})=\max \left(  S_{0x}e^{\bar{x}_{T}}-K,0\right)  $. The
calculation, though rather cumbersome, is essentially the same as for the
Asian options in section \ref{average strike option}. The integral over
$x_{T}$ is a Gaussian integral, and the remaining two integrals can be
transformed into a standard bivariate cumulative normal distribution, defined
by:%
\begin{equation}
N[a,b;\chi]=\frac{1}{2\pi \sqrt{1-\chi^{2}}}\int_{-\infty}^{a}\int_{-\infty
}^{b}\exp \left(  -\frac{1}{2\left(  1-\chi^{2}\right)  }\left(  x^{2}%
+y^{2}-2\chi xy\right)  \right)  dxdy \label{cumnormdistr}%
\end{equation}
This eventually leads to the following pricing formula for an Asian option
with a barrier:%
\begin{align}
V_{0}^{AB}  &  =e^{-rT}\left[  S_{0x}e^{\frac{T}{2}\left(  \mu-\frac
{\sigma^{2}}{6}\right)  }N\left(  d_{1},d_{2},-\sqrt{\frac{3}{4}}\rho \right)
-K~N\left(  d_{3},d_{4},-\sqrt{\frac{3}{4}}\rho \right)  \right. \nonumber \\
&  -S_{0x}~e^{\frac{3T}{\sigma^{2}}\left(  \frac{x_{S}}{T}+\frac{\sigma^{2}%
}{6}\right)  \left[  2\frac{x_{S}}{T}+\left(  \mu-\frac{\sigma^{2}}{6}\right)
\right]  }\left(  \frac{B}{S_{0y}}\right)  ^{\frac{2\left[  \frac{4}{\xi
}\left(  \nu-\frac{\xi^{2}}{2}\right)  -3\frac{\rho}{\sigma}\left(  \mu
-\frac{\sigma^{2}}{2}\right)  \right]  }{\xi \left(  4-3\rho^{2}\right)  }%
}N\left(  d_{5},d_{6},-\sqrt{\frac{3}{4}}\rho \right) \nonumber \\
&  \left.  +K~e^{\frac{3}{\sigma^{2}}x_{S}\left[  \frac{2x_{S}}{T}+\left(
\mu-\frac{\sigma^{2}}{2}\right)  \right]  }\left(  \frac{B}{S_{0y}}\right)
^{\frac{2\left[  \frac{4}{\xi}\left(  \nu-\frac{\xi^{2}}{2}\right)
-3\frac{\rho}{\sigma}\left(  \mu-\frac{\sigma^{2}}{2}\right)  \right]  }%
{\xi \left(  4-3\rho^{2}\right)  }}N\left(  d_{7},d_{8},-\sqrt{\frac{3}{4}}%
\rho \right)  \right]  \label{Optieprijs aziaat met barriere}%
\end{align}
where the following shorthand notations were used:%
\begin{equation}%
\begin{array}
[c]{l}%
d_{1}=-\dfrac{\ln \left(  \dfrac{K}{S_{0x}}\right)  -\dfrac{T}{2}\left(
\mu+\dfrac{\sigma^{2}}{6}\right)  }{\sqrt{\dfrac{\sigma^{2}T}{3}}}\\
d_{2}=\dfrac{\ln \left(  \dfrac{B}{S_{0y}}\right)  -T\left(  \nu-\dfrac{\xi
^{2}}{2}+\dfrac{\sigma \xi \rho}{2}\right)  }{\sqrt{\xi^{2}T}}\\
d_{3}=-\dfrac{\ln \left(  \dfrac{K}{S_{0x}}\right)  -\dfrac{T}{2}\left(
\mu-\dfrac{\sigma^{2}}{2}\right)  }{\sqrt{\dfrac{\sigma^{2}T}{3}}}\\
d_{4}=\dfrac{\ln \left(  \dfrac{B}{S_{0y}}\right)  -T\left(  \nu-\dfrac{\xi
^{2}}{2}\right)  }{\sqrt{\xi^{2}T}}\\
d_{5}=-\dfrac{\ln \left(  \dfrac{K}{S_{0x}}\right)  -T\left[  2\dfrac{x_{S}}%
{T}+\dfrac{1}{2}\left(  \mu+\dfrac{\sigma^{2}}{6}\right)  \right]  }%
{\sqrt{\dfrac{\sigma^{2}T}{3}}}\\
d_{6}=\dfrac{\ln \left(  \dfrac{B}{S_{0y}}\right)  -\dfrac{T}{\sigma}\left[
3\xi \rho \dfrac{x_{S}}{T}+\sigma \dfrac{x_{S}}{T}+\sigma \left(  \nu-\dfrac
{\xi^{2}}{2}+\dfrac{\sigma \xi \rho}{2}\right)  \right]  }{\sqrt{\xi^{2}T}}\\
d_{7}=-\dfrac{\ln \left(  \dfrac{K}{S_{0x}}\right)  -T\left[  \dfrac{2x_{S}}%
{T}+\dfrac{1}{2}\left(  \mu-\dfrac{\sigma^{2}}{2}\right)  \right]  }%
{\sqrt{\dfrac{\sigma^{2}T}{3}}}\\
d_{8}=\dfrac{\ln \left(  \dfrac{B}{S_{0y}}\right)  -\dfrac{T}{2\sigma}\left[
\dfrac{1}{T}\left(  6\rho x_{S}\xi+2\sigma y_{S}\right)  +2\sigma \left(
\nu-\dfrac{\xi^{2}}{2}\right)  \right]  }{\sqrt{\xi^{2}T}}%
\end{array}
\end{equation}

\subsection{Results and discussion}

\begin{figure}
[h]
\begin{center}
\includegraphics[
height=3.5583in,
width=5.6134in
]%
{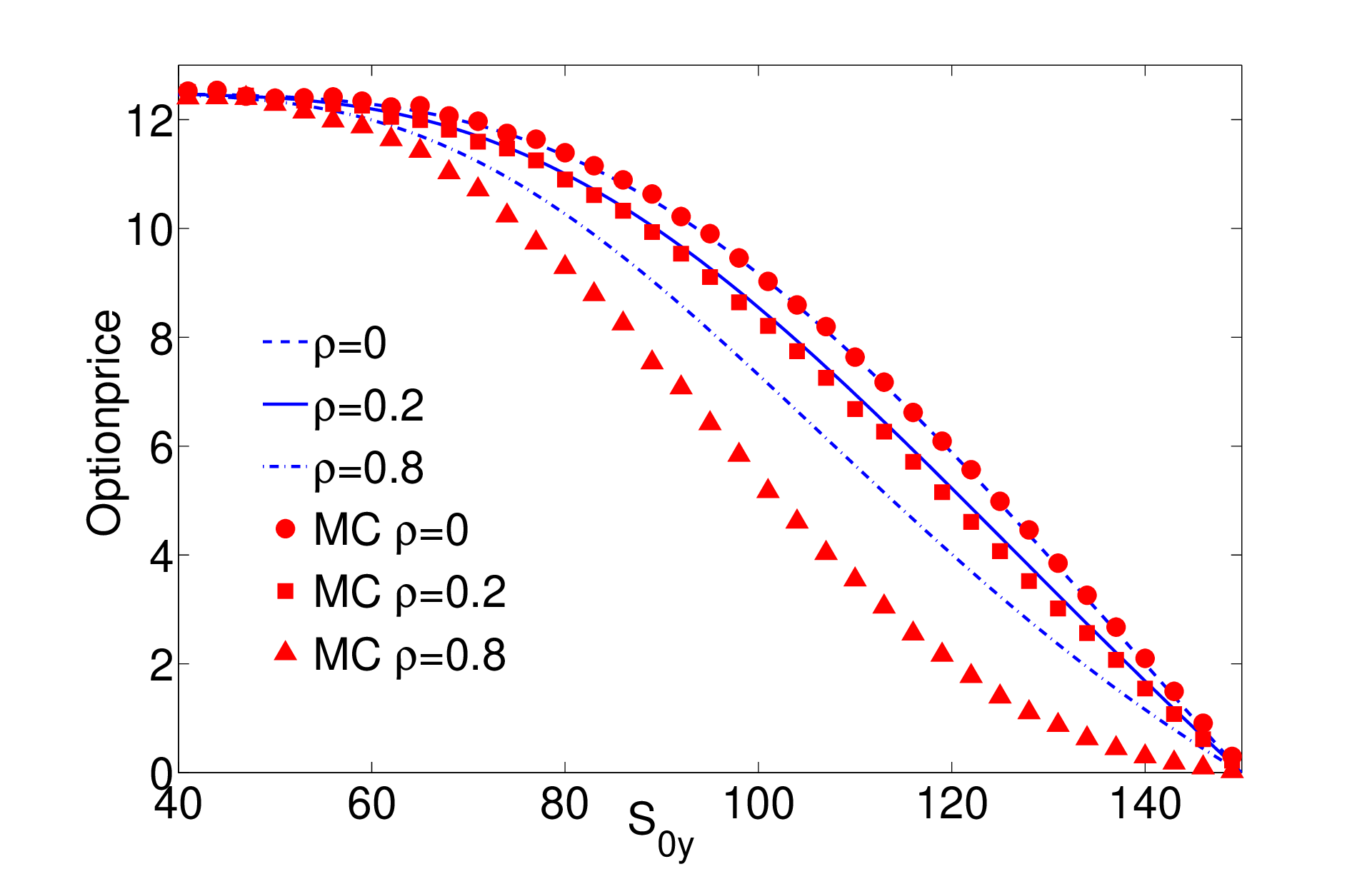}%
\caption{Option price for an Asian option with a barrier, as a function of the
initial asset price belonging to the $y$ process. Since no specific asset is
considered the option price could be stated in any currency, therefore this
figure is given in arbitrary units. The analytical result deviates from the
Monte Carlo simulation (MC) for increasing correlation. The value for the
barrier used in this figure is $B=150\text{.}$}%
\label{versch_corr_artikel}%
\end{center}
\end{figure}

Fig. (\ref{versch_corr_artikel}) shows the option price for an Asian option
with a barrier as a function of the initial asset price belonging to the $y$
process, defined by: $S_{Ty}=S_{0y}e^{y_{T}}$.
This figure shows that the analytical result derived in section
\ref{derivation of the option price} deviates from the Monte Carlo simulation
with increasing correlation. The approximate nature of our approach can be
understood as follows. The essence of the approach presented here is that to
calculate the price of Asian barrier options, two steps need to be taken.
First, a partitioning of paths according to the average along the path must be
performed, and second, the method of images must be used in order to cancel
out paths which have reached the barrier. The difficulty combining these two
steps, is that mirror paths have a different average than the original paths,
and thus belong to a different partition. This difficulty can apparently be
overcome by treating the average itself as a separate, correlated process (as
proposed in Ref. \cite{Glassy}). This procedure, relating $\mathcal{K}\left(
x_{T},T\, \left \vert 0,0\right \vert \bar{x}_{T}\right)  $ to $\mathcal{K}%
\left(  x_{T},\bar{x}_{T},T\,|0,0,0\right)  $, leads to the correct propagator
(and price) in the case of a plain Asian option as shown in section \ref{2}.

However, from the results shown in Fig. \ref{versch_corr_artikel} it is clear
that this is no longer the case for an Asian option with a barrier on a
correlated control process. This is because the exact average of the $x$
process does not behave as a separate, correlated process (the average
described by this process is henceforth called the approximate average). This
approach is exact for a plain Asian option, where all paths contribute, but
when a barrier is implemented using the method of images, and thus eliminating
some of the paths, the following approximation is made. When the $y$ process
hits the barrier and is thus eliminated, its corresponding $x$ and $\bar{x}$
processes are eliminated as well. But the $\bar{x}$ process considered in our
derivation is only approximate, so the wrong $\bar{x}$ paths are eliminated.
The central question is whether this will lead to a difference between the
distribution of contributing paths for the exact averages and the
corresponding distribution for the approximate averages, when a barrier has
been implemented. Figure 3 shows that this is indeed the case, and that this
difference increases when correlation increases.%
\begin{figure}
[h]
\begin{center}
\includegraphics[
height=3.7505in,
width=6.5461in
]%
{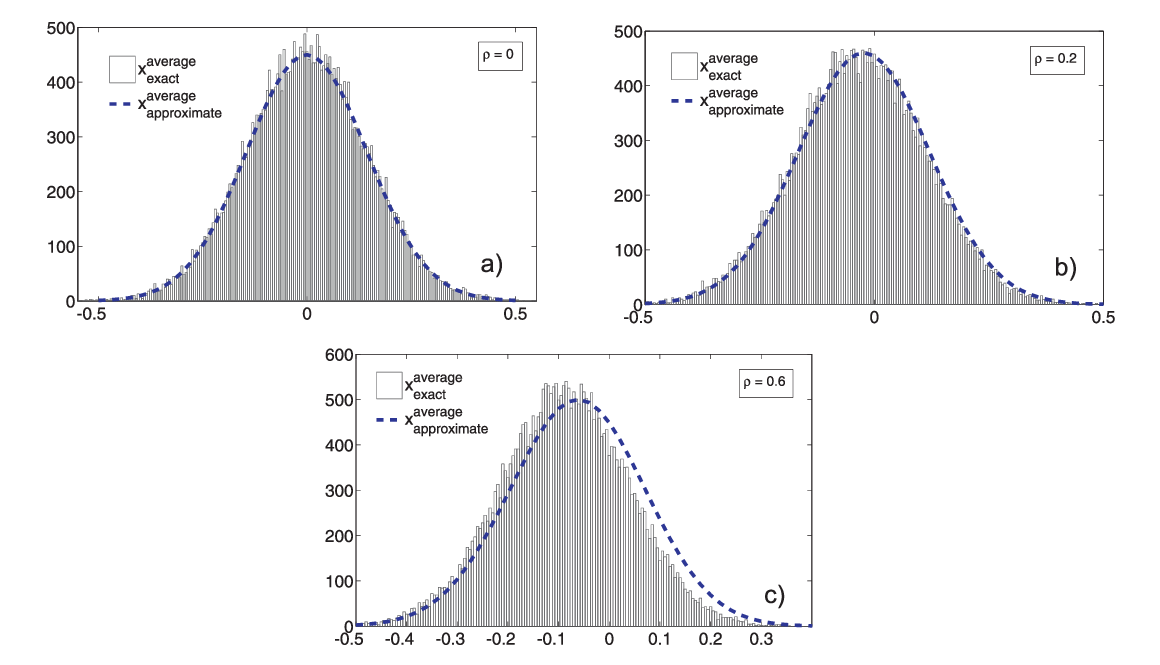}%
\caption{Comparison of the distribution of the exact and the approximate
average of the logreturn\protect \nolinebreak \ $x$, when a barrier is present.
a) When the correlation equals zero, both distributions are equal b) For
relative small values of the correlation, the assumption that the $\bar{x}$
process is correlated with the process for the logreturn $x$ is still
relatively good. \ c) When the correlation is relatively large, the
distribution of the exact averages deviates significantly from the approximate
averages.}%
\label{histogrammen_bijeen_artikel}%
\end{center}
\end{figure}
When the correlation is zero, the paths which are eliminated for both the
exact and the approximate average are randomly distributed (because the
behavior of $\bar{x}$ has nothing to do with the behavior of $y$) , which
means that both distributions remain the same Gaussian as they would be
without a barrier. This is the reason why our result is exact when correlation
is zero. Another source of approximation lies in the use of the Black-Scholes
model which has well-known limitations \cite{Bouchaud,Hull}. Several other
types of market models propose to overcome such limitations, for example by
introducing additional ad hoc stochastic variables \cite{Heston} or by
improving the description of the behavior of buyers/sellers \cite{Medo}. The
extension of the present work to for example the Heston model lies beyond the
scope of this article.

\section{Conclusions\label{5}}

In this paper, we derived a closed-form pricing formula for an average price
as well as an average strike geometric Asian option within the path integral
framework. The result for the average price Asian option corresponds to that
found by Linetsky \cite{Linetsky}, using the effective classical partition
function technique developed by Feynman and Kleinert \cite{Feynman-Kleinert}.
The result for the average strike Asian option was compared to a Monte Carlo
simulation. We found that the agreement between the numerical simulation and
the analytical result for an average strike Asian option is such that they
coincide to within a relative error of less than 0.3 \% for at least 500 000
samples and 100 time steps.

Furthermore, a pricing formula for an Asian option with a barrier on a control
process was developed. This is an Asian option with the additional condition
that the payoff is zero whenever the value of $\allowbreak$the control process
crosses a certain predetermined barrier. The pricing of this option was
performed by constructing a new propagator which consisted of a linear
combination of two propagators for a regular Asian option. The resulting
pricing formula is exact when the correlation is zero, and is approximate when
the correlation increases. The central approximation made in our derivation,
is that the process for the average logreturn $\bar{x}$ is treated as a
stochastic process, which is correlated with the process of the logreturn $x$.
This assumption is correct whenever all price-paths contribute to the total
sum, but becomes approximate when a boundary condition is applied.

\begin{acknowledgments}
The authors would like to thank Dr. Sven Foulon and prof. dr. Karel in 't Hout
for the fruitful discussions. This work is supported financially by the Fund
for Scientific Research-Flanders, FWO project G.0125.08, and by the Special
Research Fund of the University of Antwerp BOF NOI UA 2007.
\end{acknowledgments}

\end{document}